# Defect engineering for control of wake-up effect in $HfO_2$-based ferroelectrics

Alireza Kashir*, Seungyeol Oh, Hyunsang Hwang**

Center for Single Atom–based Semiconductor Device and Department of Materials Science and Engineering, Pohang University of Science and Technology (POSTECH), Pohang, Republic of Korea

E-mail:

*kashir@postech.ac.kr

**hwanghs@postech.ac.kr

Wake-up effect is still an obstacle in the commercialization of hafnia-based ferroelectric thin films. In this work, we investigate the effect of defects, controlled by ozone dosage, on the field cycling behavior of the atomic layer deposited $Hf_{0.5}Zr_{0.5}O_2$ (HZO) films. A nearly wake-up free device was achieved after reduction of carbon contamination and oxygen defects by increasing the ozone dosage. The sample which was grown at 30 sec ozone pulse duration shows about 98% of the woken-up $P_r$ at the pristine state while those grown below 5 sec ozone pulse time show a pinched hysteresis loop, undergone a large wake-up effect. This behavior is attributed to the increase in oxygen vacancy and carbon concentration in the films deposited at insufficient $O_3$ dosage which was confirmed by x-ray photoelectron spectroscopy (XPS). X-ray diffraction (XRD) scan shows that the increase of ozone pulse time yields in the reduction of tetragonal phase; therefore, the dielectric constant reduces. The I-V measurements reveal the increase of current density as the ozone dosage decreases which might be due to the generation of oxygen vacancies in the deposited film. Finally, we have investigated the dynamics of wake-up effect and it appears to be explained well by Johnson-Mehl-Avrami-Kolmogoroff model which is based on structural phase transformation.

**Keywords:** Wake-up effect; $HfO_2$; Defect engineering; Ferroelectrics; Carbon content; X-ray photoelectron spectroscopy; Johnson-Mehl-Avrami-Kolmogoroff model; Leakage current;

## Introduction

Ferroelectricity in $HfO_2$-based thin films was observed first time in 2011 by Boscke et al. [1] and thereafter, due to its numerous advantages such as simple structure, strong binding energy between the oxygen and transition metal ions, wide bandgap (~ 5.3-5.7 eV) and more importantly, the compatibility with current complementary metal oxide semiconductor (CMOS) technologies, an extensive research has been conducted to achieve a reliable property for various potential applications such as ferroelectric memory, ferroelectric field effect transistors (Fe-FETs), energy harvesters, pyroelectric sensors and so on.

Bulk $HfO_2$ crystalizes in monoclinic ($P2_1/c$) structure (m-phase) at ambient condition [2]. A Martensitic phase transformation from monoclinic to tetragonal ($P4_2/nmc$) (t-phase) is observed at ~ 1700 ºC. A further increase in temperature to ~ 2200 ºC causes an diffusionless tetragonal to cubic $Fm\bar{3}m$ (c-phase) transition [2]. These transition temperatures can be substantially altered by doping, mechanical stress or surface manipulation and the high temperature phases are practically achieved in $HfO_2$ thin films at the room temperature (RT) [3-14]. A non-centrosymmetric polar orthorhombic phase ($Pca2_1$, o-phase) is believed to be the structural origin of ferroelectricity in $HfO_2$ based thin films [2, 15]. $Pca2_1$ is extremely close in free energy ($< k_BT/5$, where $k_B$ is the Boltzmann constant) to the equilibrium nonpolar phases [2]. Therefore, tuning experimental conditions (e.g., via strain or dopants) may stabilize this polar polymorph. In fact, the transition of t-phase to o-phase instead of m-phase, is a response to the large volume expansion of crystal by ~ 5% during the later transition under the capping layer confinement. The t- to o-phase transition causes ~ 1% decrease in volume [8]. Barta et al. [5] predicted a transition from non-polar to polar $Pca2_1$ phase by applying a strong electric field combined with the application of an appropriate strain. Therefore, electric field can provide an additional driving force to induce ferroelectricity in $HfO_2$-based thin films.

Despite many researches on $HfO_2$ based ferroelectric thin films and obtaining strong ferroelectric properties, within the lifetime of the device, two critical problems *i.e.* wake-up and fatigue can be identified which cause serious obstacles for device operation. It was experimentally observed that wake-up effect is due to the migration of point defects (mainly oxygen vacancies) and a transition from non-ferroelectric (m- and t-phase) to ferroelectric phase (especially at the interfacial region).

Both mechanisms can be driven under application of a strong cycling electric field. A locally distributed inhomogeneous internal field originated from the spatially unevenly distributed charged defects, such as oxygen vacancies can be an origin of the internal field in the pristine material [16-21]. During the electric field cycling process, the oxygen vacancies may diffuse into the bulk regions of the ferroelectric-$HfO_2$-based films, in which case the wake-up process will occur. Starschich et al. [21] showed that even a DC pulse with an adequate length wakes up the doped $HfO_2$ capacitors. Thus, the total pulse length may be critical for the wake-up effect, implying a time-dependent phenomenon. Another mechanism which is responsible for wake-up effect is the structural transition from non-ferroelectric (m or t) to ferroelectric (o) phase driven by a strong electric field [22-23]. Therefore, wake-up effect can be substantially reduced by reduction of m- and t-phase during the thin film processing [24]. Goh et al. [25] evaluated the effect of metal oxide (as an electrode) on the wake-up behavior of the $Hf_{0.5}Zr_{0.5}O_2$ (HZO) film and they observed an improve in wake-up behavior which could be due to the suppression of oxygen vacancy at the interfacial region. Oxide metal electrode can provide additional oxygen to the HZO layer to hinder the oxygen vacancy generation during the application of electric field. Therefore, reduction of oxygen vacancies (in both bulk and interfacial regions) and suppression of non-ferroelectric (m and t) phases during $HfO_2$-based thin film processing, are the promising paths towards wake-up free devices.

The ozone $O_3$ pulses during the atomic layer deposition (ALD) of $HfO_2$-based thin film have two major roles [26]. The primary role is to infuse oxygen into Hf and Zr layers to form $HfO_2$ and $ZrO_2$ compounds. Thus, the pulse length is a crucial parameter to provide enough oxygen for stoichiometric phase formation and avoiding the generation of oxygen vacancies. But it should be noticed that the C−O bonds removal out of the sample by applying $O_3$ pulses can remarkably change the subsequent annealing behavior of the as-grown film [26]. Carbon is an inevitable part of the ALD grown films as the precursors are usually organic compounds. Cho et al. [26] observed the appearance of tetragonal phase as the ozone dosage decreases which was attributed to the effect of carbonate bonds on the annealing behavior of $HfO_2$ films. The C-O bonds remain between pure $HfO_2$ domains due to the incomplete chemical reaction of the precursors which subsequently prevent the agglomeration of the nanoscale domains which result in the stabilization of tetragonal phase [26-28]. Therefore, ozone dosage can drive the wake-up effect directly by oxygen vacancy generation and indirectly by carbon-induce t-phase formation.

As a proper capping electrode, tungsten (W) shows the lowest thermal expansion coefficient $\alpha$ among the widely used metal electrodes, inducing an in-plane tensile strain to HZO film during the rapid thermal annealing (RTA), which facilitates the formation of o-phase [29]. In-plane tensile strain stresses on c-axis of t-phase in favor of the phase change from t- to o-phase. Moreover, $WO_2$ and $WO_3$ formation enthalpy is almost two times higher than that of $HfO_2$ and $ZrO_2$, preventing the formation of interfacial oxide during atomic layer deposition of HZO which in turn facilitates the deposition of stoichiometric $Hf_{0.5}Zr_{0.5}O_2$ by an appropriate tuning of ozone dosage.

Considering these facts, we investigate the effect of $O_3$ pulse time (ozone dosage) during atomic layer deposition of the ~ 10 nm HZO films, on the wake-up behavior of W/HZO/W capacitors. The both major roles of oxygen will be considered. We will precisely study the effect of remaining carbon and $O_3$ pulse length on the ferroelectric behavior of the HZO thin films. Finally, the kinetics of wake-up effect is explained by Johnson-Mehl-Avrami-Kolmogoroff model.

## Experiments

The ~ 10 nm HZO films were deposited on 50-nm thick W bottom electrode sputtered on $SiO_2$/Si substrate using ALD technique. Different ozone pulse durations $t_{O_3}$= 2 to 30 sec were applied for each deposition. The substrate temperature was kept at 250 ºC during the deposition of all films in this work. The $Hf[N-(C_2H_5)CH_3]_4$ and $Zr[N-(C_2H_5)CH_3]_4$ precursors were used as the Hf and Zr metals sources, respectively. The growth rate of $HfO_2$ and $ZrO_2$ were almost same (~ 1 Å/cycle). After deposition, all films were capped with 50-nm W electrode using rf-sputtering technique to fabricate MIM capacitors. Finally, W/HZO/W capacitors with different electrode area were passed through an annealing process in $N_2$ ambient at 500 ºC for 30 sec.

The crystalline structures of films were investigated using an x-ray diffractometer (XRD) within a grazing incidence geometry. For elemental analysis we used x-ray photoelectron spectroscopy (XPS) on as-grown samples (not capped). The ferroelectric properties of MIM capacitors were measured by LCII ferroelectric precision tester (Radiant Technologies) and the dielectric permittivity and I–V characteristics were evaluated using Keysight B1500A semiconductor device parameter analyzer. All measurements were conducted at room temperature.

## Results and Discussion

Figure 1 shows the pristine P-E hysteresis curves for the samples grown at different $t_{O_3}$. Increasing $t_{O_3}$ results in a more open (depinched) loop. Therefore, the pristine $2P_r$ value increases as the samples are grown at longer ozone pulse length. The P-E curves at the wake-up state for two samples deposited at $t_{O_3}$=5 and 30 sec are shown in figure 2. It revealed that the sample which was grown at lower $t_{O_3}$ undergoes a large wake-up effect, while the sample grown at higher $t_{O_3}$ is almost wake-up free and its pristine $2P_r$ value is about 97% of the wake-up state (Table 1). This is only 63% in case of the sample grown at $t_{O_3}$=5 sec. Lee *et al.* [19] using Monte Carlo simulation demonstrated that the ferroelectric hysteresis of a doped $HfO_2$ thin film with a lower defect ratio exhibits an increase in both $P_r$ and the coercive field $E_c$. Defects can pin the ferroelectric domains and strongly affect the ferroelectric properties of thin film. As we discussed, two major sources which lead to the wake-up effect were recognized; the charged oxygen vacancies and the structural phase transition from non-ferroelectric to ferroelectric phase. On the other hand, the oxygen vacancy movement during the field cycling causes a local phase transition in different region, as the stability of different polymorphs in $HfO_2$-based thin films depends on the amount of oxygen vacancy [30-31]. In a recent work, based on the DFT calculation, it was predicted that even m-phase which is originally non-ferroelectric phase shows ferroelectricity by introduction of oxygen vacancy [32]. Therefore, the oxygen vacancy and its movement during electric field cycling has one of the major contributions in wake-up effect (either by pinning of ferroelectric domain walls or by stabilizing o-phase). The decrease of $O_3$ dosage during the deposition of HZO might induce point defects, dominantly oxygen vacancies, and subsequently trigger the mechanisms of wake-up during the electric field cycling.

Figure 3 revealed the effect of $O_3$ dosage on the leakage current of HZO samples. At 1 MV/cm, the sample which was grown at 5 sec $O_3$ pulse shows the electric current density of ~ 2.5 times higher than that of grown at 30 sec. This behavior implies the role of ozone dosage in the reduction of defects and subsequently improving the insulating characteristics of HZO films.

It should be noted that the removing of C−O bonds out of the sample by applying $O_3$ pulse can remarkably influences the subsequent annealing behavior of the as-grown film [26-28]. This behavior becomes much more important and crucial when the ferroelectricity of $HfO_2$-based thin film is the major concern. Figure 4 shows the XPS spectra of the samples grown at two different $O_3$ pulse duration. The Zr/Hf ratio was almost 1 for both sample, indicating the $(Zr_{0.5}Hf_{0.5})O_{2-x}$ composition. In case of the sample grown at $t_{O_3}$= 5 sec, we detected smaller fraction of oxygen (~ 1%) compared to the sample which was grown at $t_{O_3}$=30 sec (Figure 4c). The sample with more oxygen deficiency shows higher amount of carbon (Figure 4d). The XPS measurement revealed around 2.1% carbon in the sample which was grown at $t_{O_3}$= 5 sec, while the sample grown at higher $O_3$ dosage were almost free of carbon. These measurements revealed the role of $O_3$ pulse time in the generation of oxygen defects and carbon contaminants in the atomic layer deposited films.

It was shown that the less $O_3$ pulse time causes higher amount of carbon in the as-grown sample because of the insufficient precursor oxidation, which subsequently influences the grain size of crystallized film. The smaller grain size would result in the t-phase formation due to its lower grain boundary energy compared to m- and even o-phases (Figure 5) [26-28]. The formation of t-phase and its transformation to o-phase under strong electric field is another cause for wake-up effect. Therefore, applying longer $O_3$ pulses can decrease the amount of oxygen defects and carbon content in the as-grown film which consequently hinder the mechanisms of wake-up effect.

XRD scans clearly demonstrate the decrease of t-phase and increase of o-phase with increasing the ozone dosage during the growth (Figure 6). For clear visualization, we showed a limited range of 2θ in which the evolution of HZO o- and t-phase Bragg peaks versus ozone dosage are clearly observable. Increasing the ozone dosage from 5 to 30 seconds results in a decrease in t-phase from 29% to 7% which considerably affect the ferroelectric properties (Table 2). On the other hand, the existence of t-phase introduces more grain boundaries which is accompanied by more defective structure. These defects can cause the wake-up effect by pinning the ferroelectric domain walls [24]. The formation of m-phase was totally suppressed during RTA under the capping electrode confinement (not sown here). All films contained t- and o-phases, possibly arising from the in-plane tensile strain applied to the HZO film during thermal annealing. Mechanical stress induced by the W-capping layer during the RTA is believed to be the driving force for the t- to o-phase transition [29]. In fact, the in-plane tensile strain, which is developed during the annealing process, substantially suppresses the formation of the m-phase by preventing the necessary twin deformations required to form the m-phase. Therefore, the TEC of the capping material substantially alters the structural properties of HZO films.

The Capacitance versus voltage measurement shows that the sample which was grown at lower $O_3$ pulse time has higher dielectric constant $\varepsilon_r$ which can be partially due to the space-charge polarization induced by defects (Figure 7a). The initial rise in permittivity with DC field is probably due to increased movement of the domain walls which become free from defects which lock them at zero-DC field [33]. From these points of view, the sample grown at lower ozone dosage contains more defects. Increasing the $t_{O_3}$ during the growth results in HZO films with lower dielectric constant $\varepsilon_r$. This is in agreement with the XRD result. It was already observed that the carbon stabilizes the tetragonal phase in $HfO_2$ base thin film and subsequently increases $\varepsilon_r$ [30-32]. The t-phase has higher $\varepsilon_r$ than m- and even o-phase in HZO films. This behavior combined with the P-E measurement (Figure 7b) suggests two possible interpretations. The samples which were grown at higher oxygen dosage contain less oxygen vacancy, therefore, less density of built-in bias electric field and finally undergo less wake–up effect. The another possible explanation is the carbon induced t-phase formation which can consequently cause the phase transition to o-phase during the field cycling. On the other hand, as it was mentioned previously, the oxygen vacancy movement provides a driving force for phase transition during the field cycling from t- to o-phase. Finally, the effect of oxygen vacancy induced ferroelectricity in m-phase should be considered. Therefore, $O_3$ pulse duration has a critical contribution to the deposition of wake-up free devices through impeding different possible mechanisms of wake-up effect.

To investigate the nature of wake-up effect, we studied the dynamics of wake-up behavior in different samples. Figure 8a shows the evolution of $P_r$ of a sample, which was grown at the lowest $O_3$ pulse duration (2 sec), under electric field cycles. The number of cycles to reach a given $P_r$ strongly depends on the applied frequency. As the frequency of the applied electric field increases, the number of cycles for achieving a given $P_r$ increases. By plotting the $P_r$ over the *duration of the* *applied field* i.e. $\frac{cycle}{frequency}$, the curves at different frequencies show the same dependence. Thus, not the amount of cycles but the *duration of the applied electrical field* is essential for the wake-up. That means that 100 cycles at 100 Hz show the same $P_r$ as 500 cycles at 500 Hz. Figure 8b revealed that the evolution of $P_r$ versus $\frac{cycle}{frequency}$ is totally independent of the applied frequency and it is a time-dependent phenomenon [21]. It can be the direct electric field induced structural change, electric filed driven oxygen vacancy movement (redistribution) or oxygen vacancy induced phase transition. All these three mechanisms are driven by cycling field and the progress of these processes are time-dependent. The increase in coercive field $E_c$ for different samples also shows the same behavior. Here, we plot the $E_c$ versus $t_{app}$ for the sample which was grown at 30 sec $O_3$ pulse time as an example (figure 8c). Considering the structural change during the electric field cycling (either through oxygen redistribution or direct electric field driven structural transformation) as the major reason for wake-up effect, we investigate the kinetics of the phase transition explained by Johnson-Mehl-Avrami-Kolmogoroff (JMAK) model [34] (equation 1)

$$f = 1 - \exp(-kt^n) \qquad \text{equation 1}$$

Where $f$ is the fraction of transferred volume, $t$ is time, $k$ is a constant and n is an integer or half-integer which is called Avrami exponent and describes the rate and geometry of nucleation and growth. Therefore

$$ln\big(-ln(1-f)\big) = ln(k) + nln(t)$$

We correlated the increase in $P_r$ to $f$ using the equation 2,

$$f = \frac{P_r - P_{Pri}}{P_W - P_{Pri}} \qquad \text{equation 2}$$

Where $P_{pri}$ and $P_W$ are the values of $P_r$ at pristine and wake-up states, respectively. Therefore,

$$f = \begin{cases} 0, & at\ pristine\ state \\ 1, & at\ wake-up\ state \end{cases}$$

Figure 9 shows a linear plot for different samples during the wake-up transition with almost same slope (~ 0.5). This behavior indicates that the phase transition during the field cycling might have the dominant role in wake-up effect. The phase transition during wake-up process can originate from different mechanisms. A strong electric field induces structural change in hafnia-based thin films. Indeed, the transition from m- and t- to o-phase were observed by different research groups [22-23], which increase the $P_r$ in $HfO_2$-based ferroelectrics. Moreover, the oxygen vacancy diffusion into different region changes the relative stability of polymorphs and consequently causes phase transition. It can stabilize the ferroelectric phase and consequently increase the remnant polarization and wake up the ferroelectric HZO.

## Conclusion

The ferroelectric behavior of ~ 10 nm ALD grown HZO thin films at different ozone pulse duration was investigated. To induce an in-plane tensile stress and achieve m-phase-free thin film, we used W electrode which has a relatively low thermal expansion coefficient. It was observed that the samples which were deposited at $t_{O_3} < 5$ sec undergo a large wake-up effect. The XRD pattern of all samples revealed the total suppression of m-phase formation during the RTA. The fraction of t-phase decreased as the $t_{O_3}$ increased which subsequently reduced the $\varepsilon_r$ of HZO films. XPS study revealed ~ 2.1% carbon in the sample grown at 5 sec $O_3$ pulse time while the sample which was grown at 30 sec $O_3$ pulse time was almost free of carbon. The carbon can stabilize tetragonal phase. The leakage current was reduced by increasing of $t_{O_3}$which could be due to the decrease of defect in HZO film. Therefore, both carbon contaminants and oxygen vacancies caused a large wake-up effect in the samples grown at insufficient ozone environment. Tuning the oxygen pulse length during the deposition of HZO, an almost wake-up free ferroelectric thin film was achieved. Finally, we investigated the kinetics of wake-up effect based on the phase transition explained by JMAK model. All samples followed the JMAK model with almost same slope. This behavior suggests the phase transition as a major cause of wake-up effect.

## Acknowledgments

This work was supported by the National Research Foundation of Korea funded by the Korea government (MSIT), Grant No. NRF-2018R1A3B1052693.

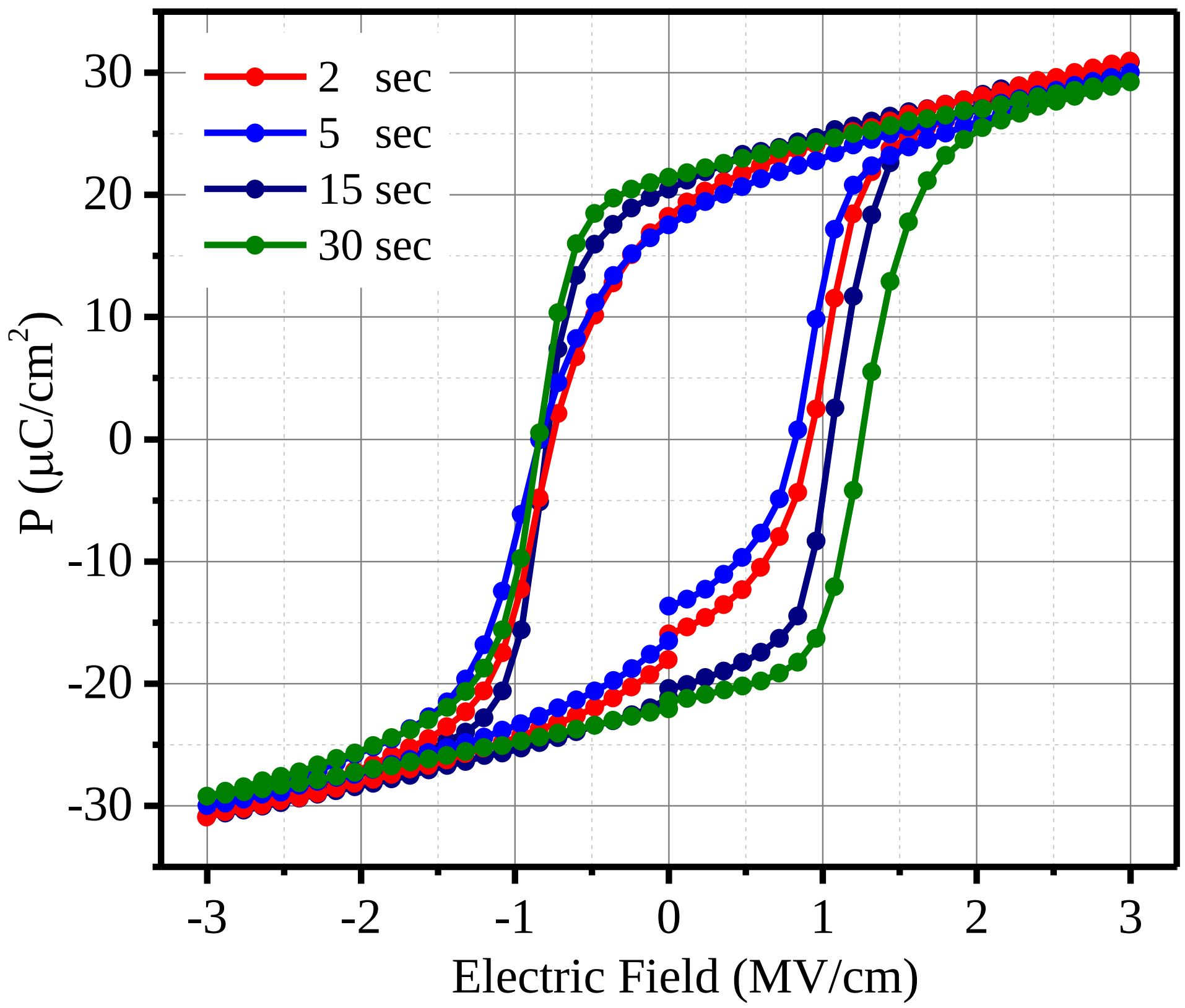

**Figure 1.** Polarization-Electric field curves of ~ 10 nm HZO films grown at different ozone pulse duration.

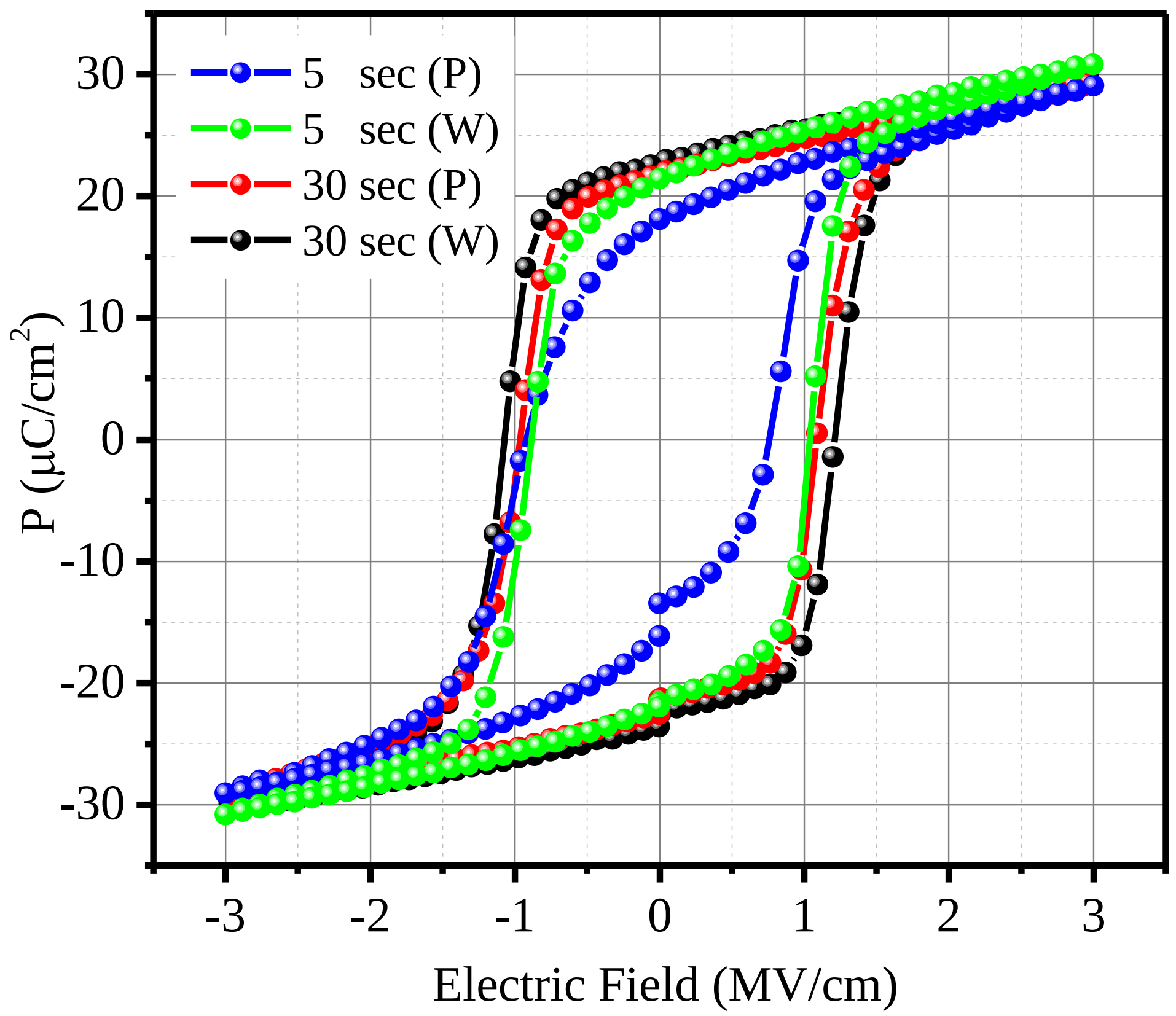

**Figure 2.** The pristine (P) and woken-up (W) Polarization-Electric field curves of ~ 10 nm HZO films grown at $O_3$ pulses.

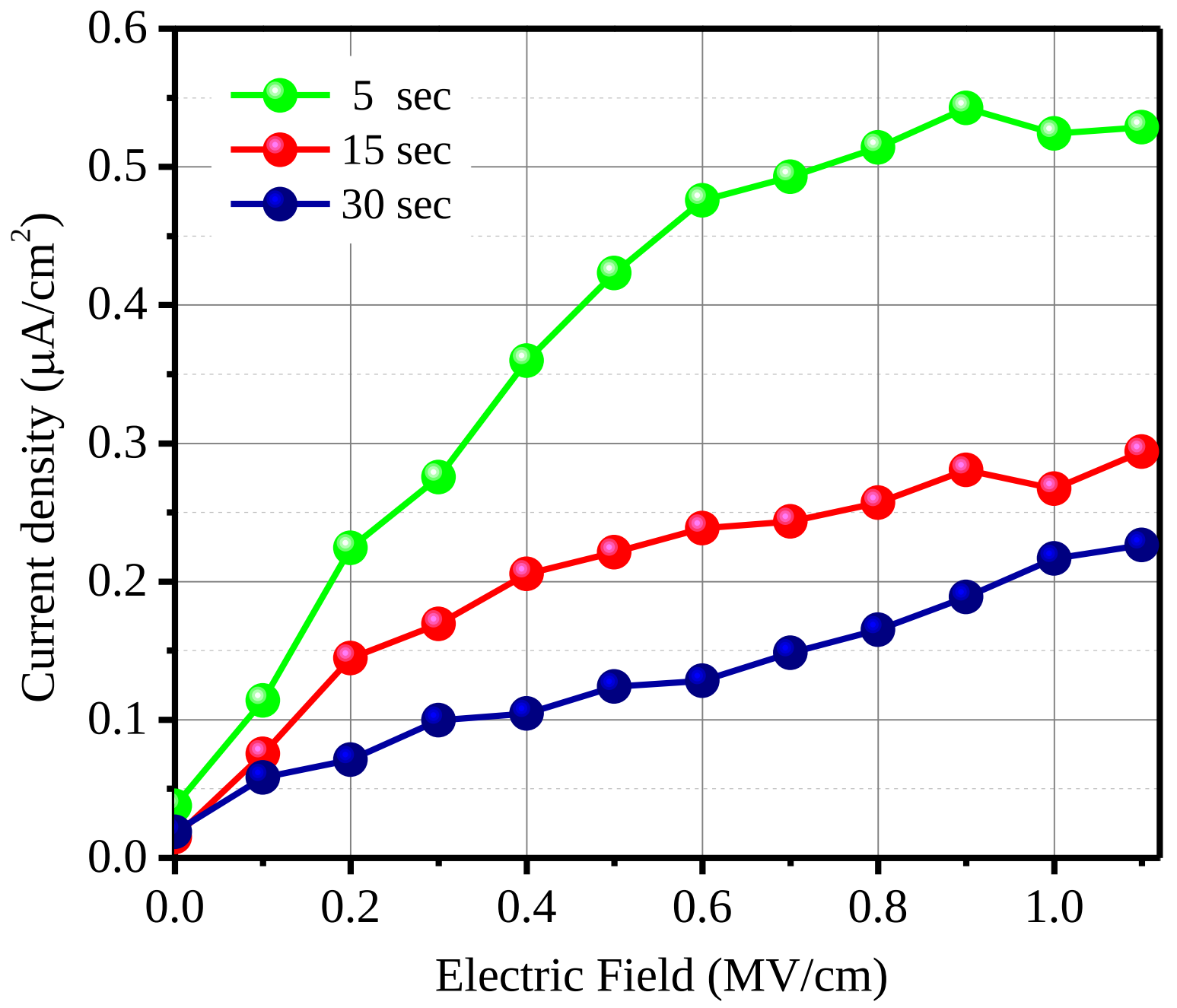

**Figure 3.** The current density vs electric field of HZO films grown at different $O_3$ pulse time.

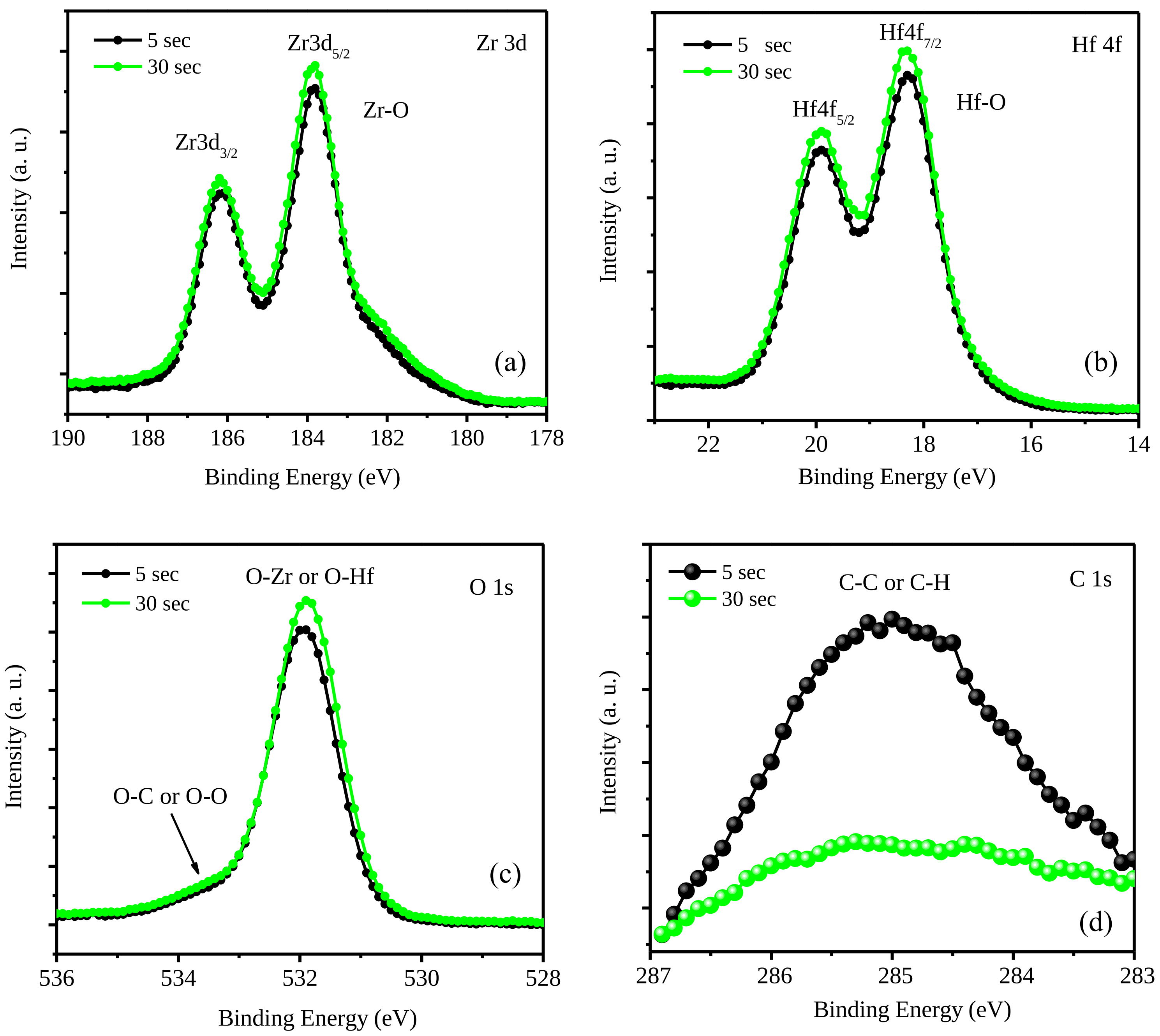

**Figure 4.** (a) Zr 3d, (b) Hf 4f, (c) O 1s, and (d) C 1s XPS spectra of the as-grown HZO thin films deposited at different $O_3$ pulse time.

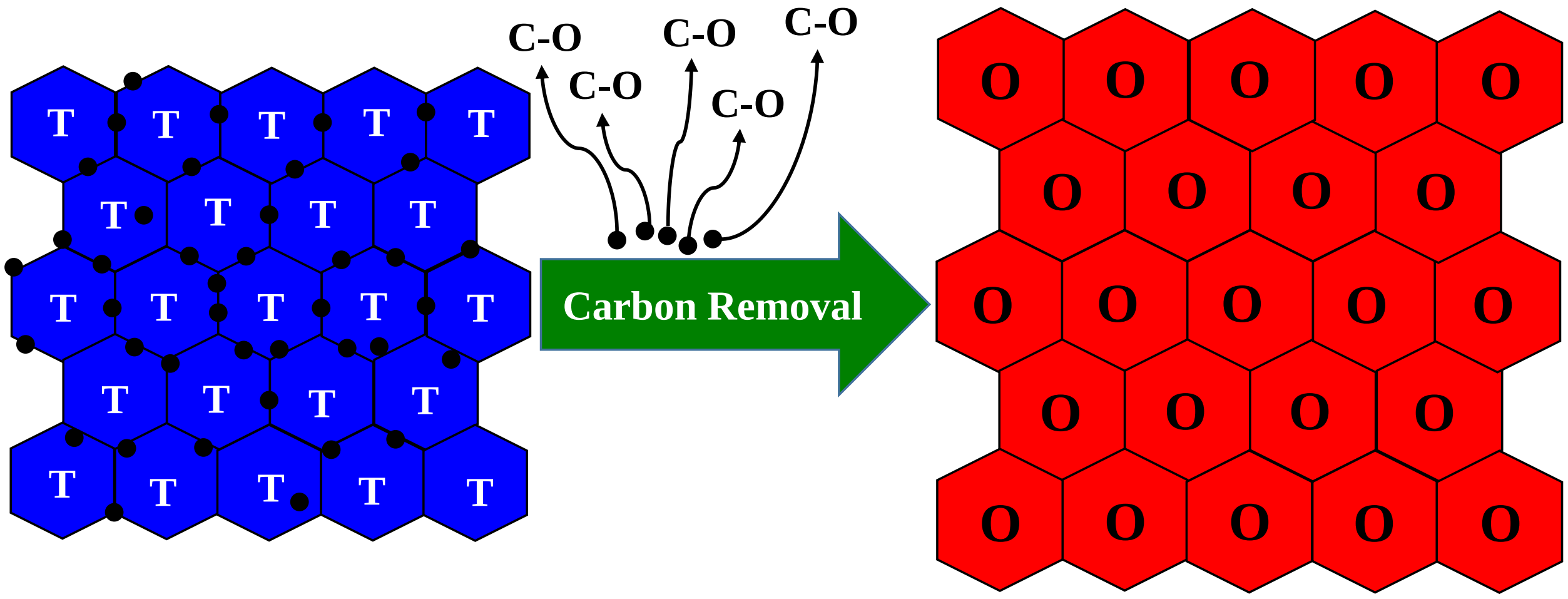

**Figure 5.** Schematic of the effect of carbon contamination (●) on the stabilization of the t-phase in a HZO thin film. Carbon removal by a sufficient oxygen supply results in the optimum growth of grain to form the ferroelectric o-phase during the cooling step of the annealing process.

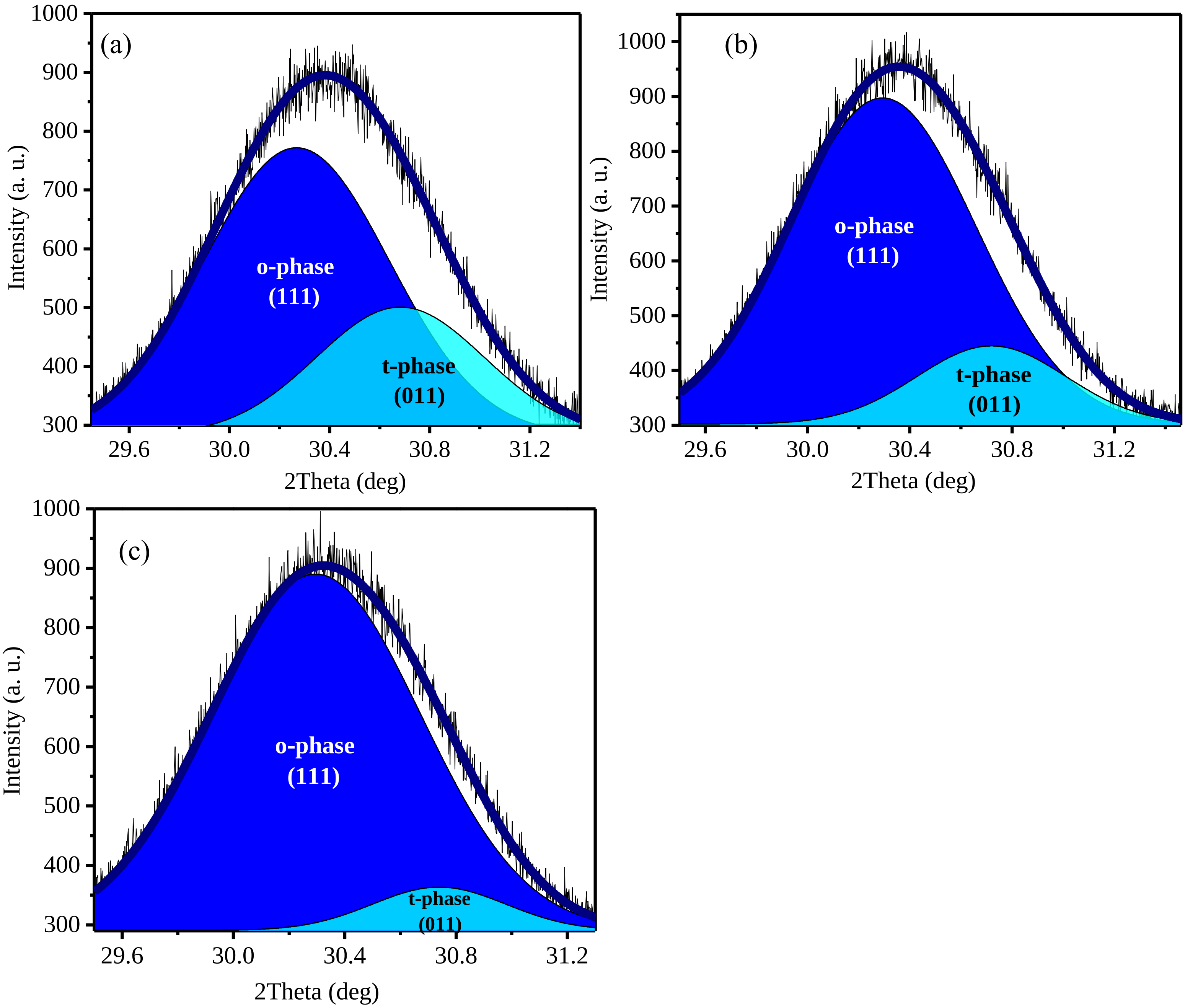

**Figure 6.** X-ray diffraction patterns of the HZO films grown at different $O_3$ pulse time, (a) 5 sec, (b) 15 sec, (c) 30 sec.

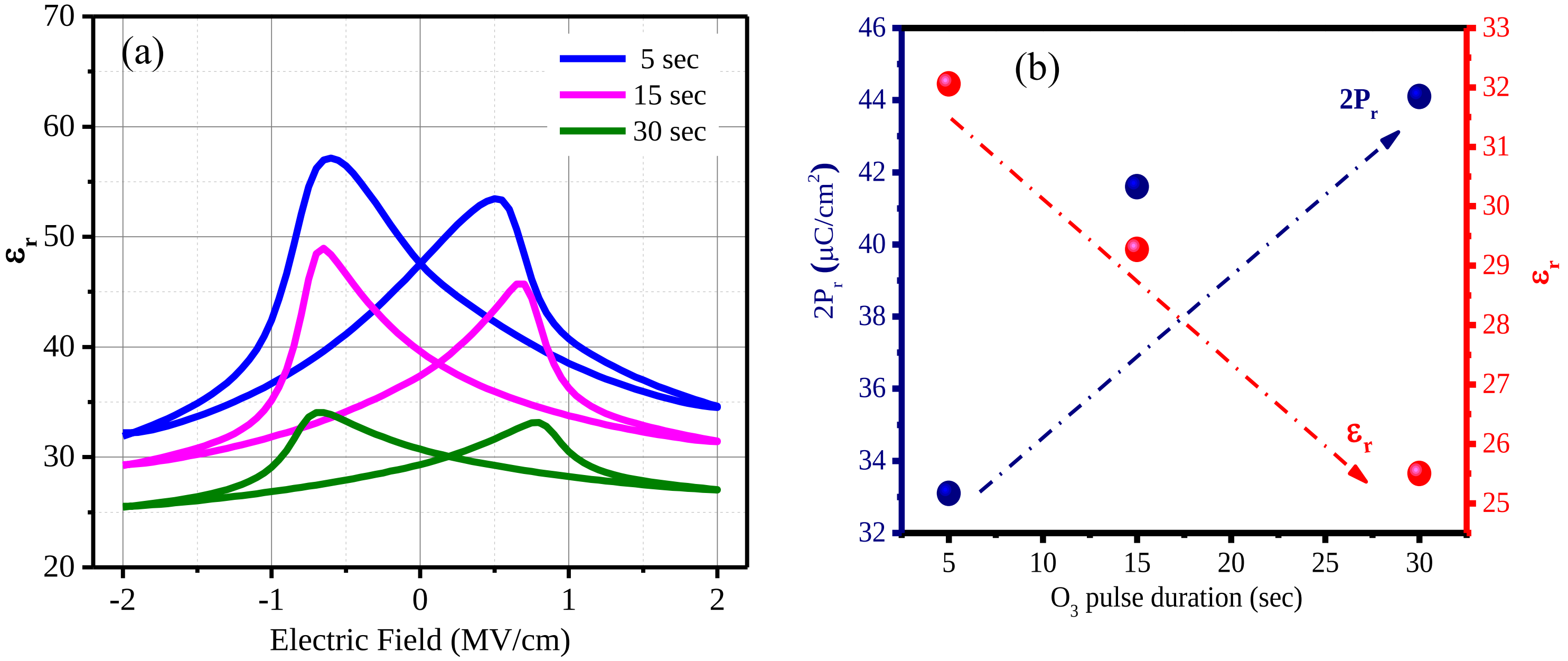

**Figure 7.** (a) The pristine dielectric constant vs electric field of the HZO thin film grown at different $O_3$ pulse time measured at 10 kHz. (b) The changes in pristine remnant polarization $2P_r$ (blue) (calculated from Figure 1) and dielectric constant $\varepsilon_r$ (red) of HZO films versus $O_3$ pulse time.

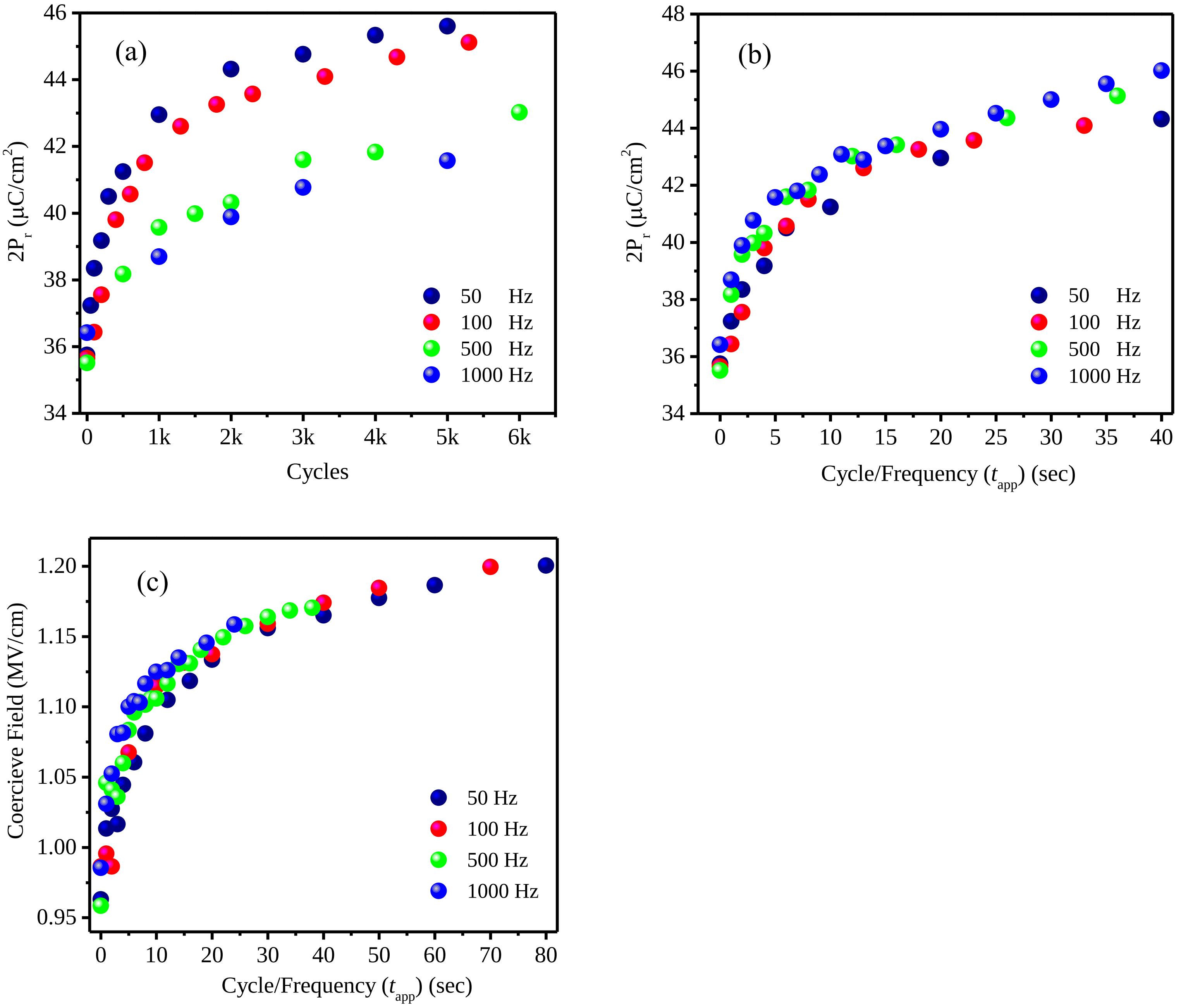

**Figure 8**. Evolution of the ferroelectric remnant polarization $P_r$ during bipolar cycling as a function of (a) cycles, (b) cycle/frequency. (c) Evolution of the ferroelectric coercive field during bipolar cycling as a function of cycle/frequency measured on HZO ferroelectric capacitor.

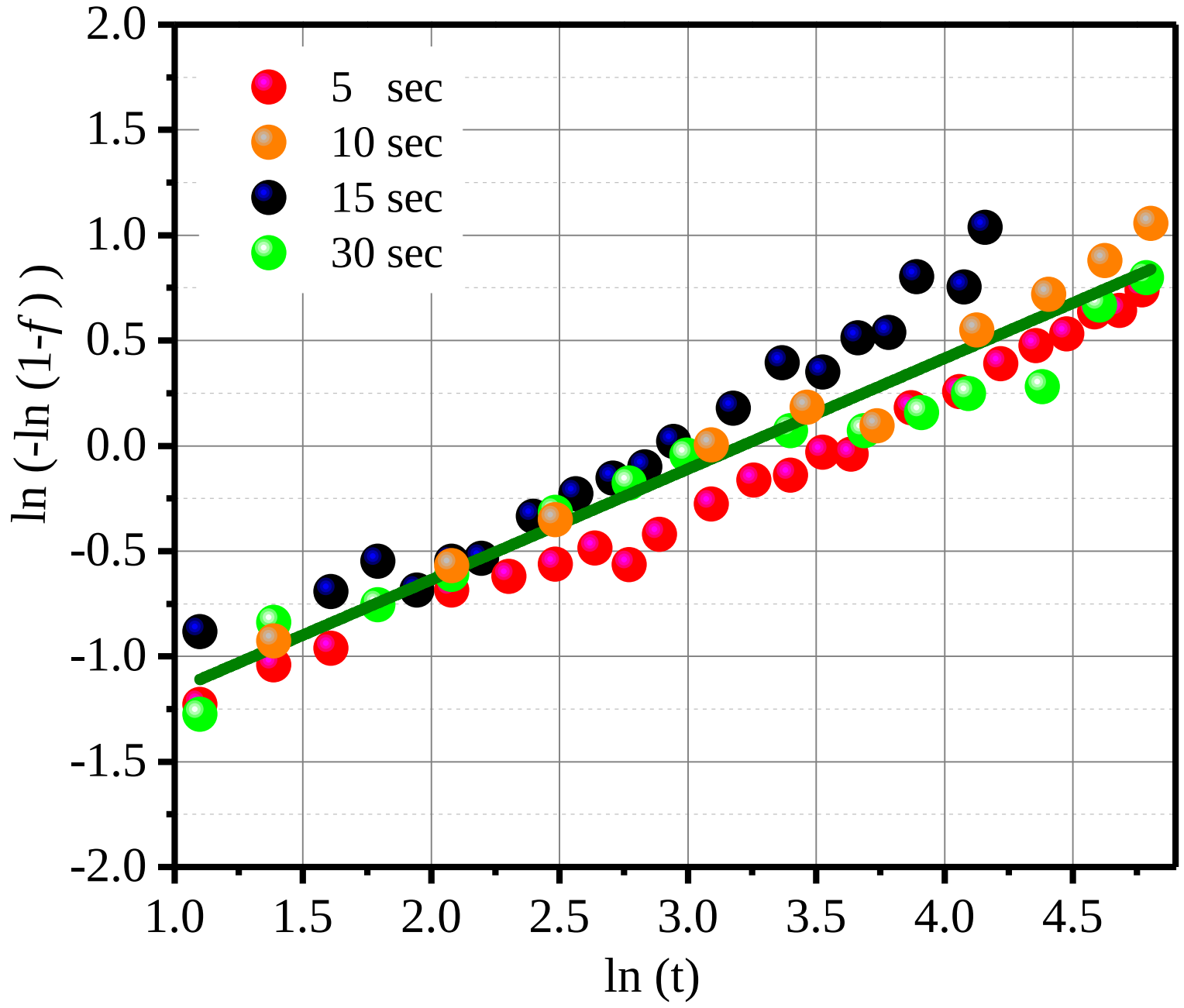

**Figure 9.** the linear plot of the fraction of transformed phase versus electric field applied time.

Table 1. The Pristine to woken-up $2P_r$ ratio of the HZO films grown at different ozone dosage.

| **Ozone pulse duration** | **2 sec** | **5 sec** | **15 sec** | **30 sec** |
|---|---|---|---|---|
| **Pristine to woken-up $2P_r$** | 63% | 63% | 90% | 97% |

Table 2. The fraction of tetragonal phase in HZO film deposited at different ozone dosage.

| **Ozone pulse duration** | **5 sec** | **15 sec** | **30 sec** |
|---|---|---|---|
| **Tetragonal phase fraction** | 29% | 16% | 7% |